\documentclass[sigconf, 9pt]{acmart}
\usepackage{algorithm}
\usepackage[noend]{algorithmic}
\usepackage{hyperref}
\usepackage{float}

\newcommand{\ydel}[1]{}
\AtBeginDocument{%
  }

\copyrightyear{2024}
\acmYear{2024}
\setcopyright{acmlicensed}\acmConference[ICCAD '24]{IEEE/ACM International Conference on Computer-Aided Design}{October 27--31, 2024}{New York, NY, USA}
\acmBooktitle{IEEE/ACM International Conference on Computer-Aided Design (ICCAD '24), October 27--31, 2024, New York, NY, USA}
\acmDOI{10.1145/3676536.3676820}
\acmISBN{979-8-4007-1077-3/24/10}

\begin{document}

\title{TSB: Tiny Shared Block for Efficient DNN Deployment on NVCIM Accelerators}

\author{Yifan Qin\textsuperscript{$\dagger\circledast$}, Zheyu Yan\textsuperscript{$\dagger$}, Zixuan Pan\textsuperscript{$\dagger$}, Wujie Wen\textsuperscript{$\ddagger$}, Xiaobo Sharon Hu\textsuperscript{$\dagger$}, Yiyu Shi\textsuperscript{$\dagger*$}}
\affiliation{%
  \institution{\textsuperscript{$\dagger$}University of Notre Dame, 
  \textsuperscript{$\ddagger$}North Carolina State University, \{\textsuperscript{$\circledast$}yqin3,
  \textsuperscript{$*$}yshi4\}@nd.edu}
  \country{}
}

\begin{abstract}
Compute-in-memory (CIM) accelerators using non-volatile memory (NVM) devices offer promising solutions for energy-efficient and low-latency Deep Neural Network (DNN) inference execution. However, practical deployment is often hindered by the challenge of dealing with the massive amount of model weight parameters impacted by the inherent device variations within non-volatile computing-in-memory (NVCIM) accelerators. This issue significantly offsets their advantages by increasing training overhead, the time and energy needed for mapping weights to device states, and diminishing inference accuracy.
To mitigate these challenges, we propose the "Tiny Shared Block (TSB)" method, which integrates a small shared $1\times1$ convolution block into the DNN architecture. This block is designed to stabilize feature processing across the network, effectively reducing the impact of device variation. Extensive experimental results show that TSB achieves over 20$\times$ inference accuracy gap improvement, over 5$\times$ training speedup, and weights-to-device mapping cost reduction while requiring less than 0.4\% of the original weights to be write-verified during programming, when compared with state-of-the-art baseline solutions.
Our approach provides a practical and efficient solution for deploying robust DNN models on NVCIM accelerators, making it a valuable contribution to the field of energy-efficient AI hardware.

\end{abstract}




\received{7 May 2024}
\received[accepted]{29 June 2024}

\maketitle

\section{Introduction}
\label{sec: intro}

\begin{figure}[t]
  \centering
  \includegraphics[scale=0.35]{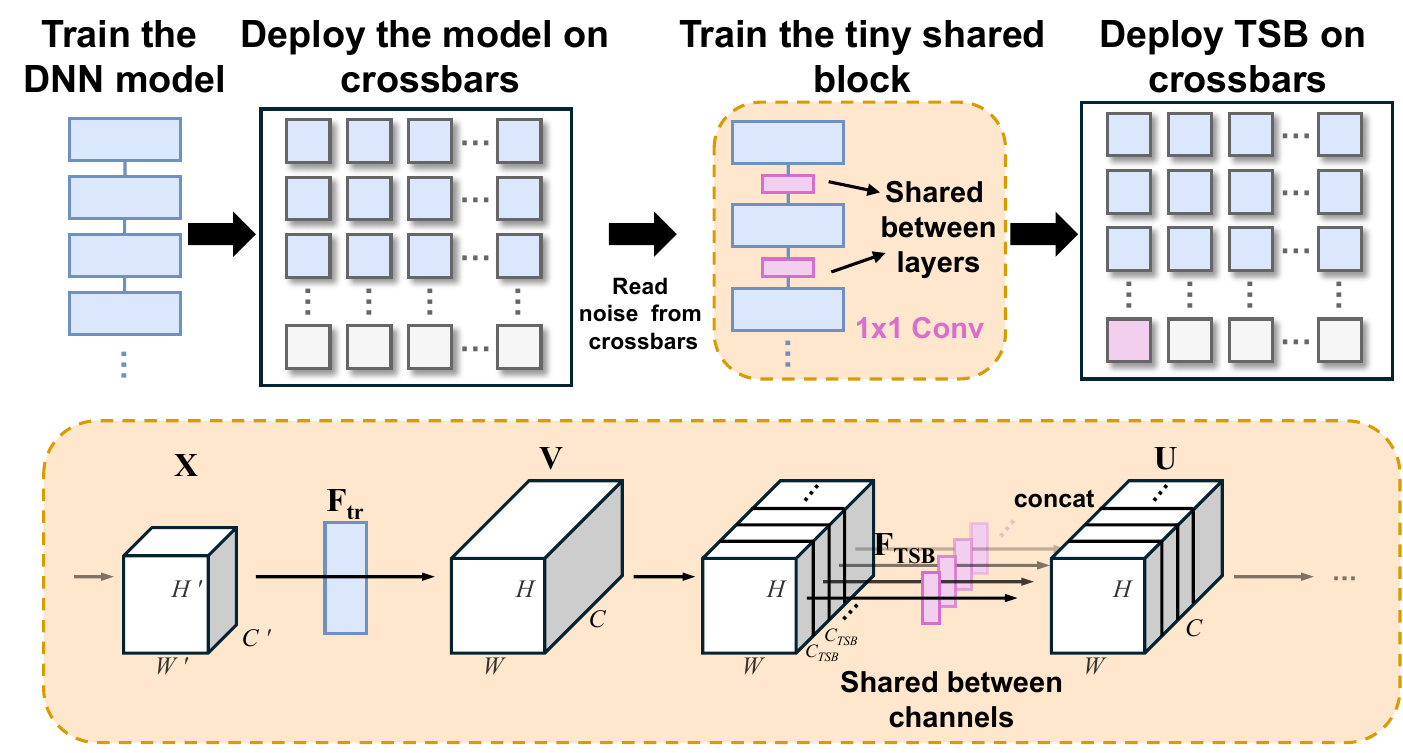}
  \caption{The tiny shared block deployment design.}
  \Description{overview}
  \label{fig: overview}
  \vspace{-15pt}
\end{figure}

Deep Neural Networks (DNNs) have sparked transformative changes across various sectors of our society. 
Nonetheless, the acceleration of DNN inference, which needs considerable vector-matrix computing, is impeded by extensive data movement between the memory and processing units \cite{liu2022efficient}.
In today's computer architecture, the varying speeds and costs of computation and memory necessitate the separation of the two \cite{xia2022pervasivefl, xia2024towards}. This separation has given rise to what is often called the ``memory wall''~\cite{mckee2004reflections} and has aggravated the von Neumann bottleneck~\cite{chen2016eyeriss}.
One promising alternative to the von Neumann architectures is computing-in-memory (CIM) based accelerators~\cite{shafiee2016isaac}. These accelerators enhance DNN inference by performing parallel in-situ vector-matrix multiplication (VMM) directly within the memory array, which is also referred to as the crossbar. Additionally, the non-volatile memory (NVM) devices, such as ferroelectric field-effect transistors (FeFETs)~\cite{reis2018computing}, resistive random-access memories (RRAMs)~\cite{qin2020design}, magnetoresistive random-access memories (MRAMs)~\cite{angizi2019mrima}, and phase-change memories (PCMs)~\cite{sun2021pcm}, on which these CIM accelerators can be built with the crossbar array further improve memory density and energy efficiency~\cite{chen2016eyeriss}. Overall, NVCIM DNN accelerators surpass the performance of conventional counterpart designs in accelerating DNN inference by significantly enhancing both time and energy efficiency~\cite{zhang2023edge}.

However, the inherent non-ideal properties of NVM devices impede the accuracy performance of NVCIM DNN accelerators. Device variation, which includes cycle-to-cycle (C2C) and device-to-device (D2D) variation~\cite{shim2020two, fritscher2022mitigating}, represents one of the most significant non-ideal factors. This variation causes conductance deviations from the targeted values~\cite{rizzi2011role, raty2015aging, chen2009highly}, and often leads to Gaussian-distributed conductance values~\cite{qin2020design} after the devices are programmed, which results in imprecise weight representations. Consequently, such device variations ultimately degrade the DNN inference accuracy on NVCIM accelerators~\cite{yan2021uncertainty}.

Achieving reliable DNN inference on unreliable NVCIM substrates, unfortunately, is a longstanding challenge in terms of \textit{accuracy maintenance, training efficiency, and deployment complexity, from the software training stage to hardware deployment}. Existing solutions, regardless of general approaches such as noise-injection training~\cite{shim2020two} or instance-specific proposals like retraining the last layer or layers~\cite{yao2020fully} and incorporating additional peripheral circuits for compensation in each accelerator instance, cannot meet these three aspects because:
\textbf{1) From the perspective of accuracy}, 
it is challenging to achieve an optimal set of weights with a high accuracy guarantee by incorporating randomly sampled noise for each individual weight in every training iteration. This difficulty arises due to the vast search space, where each parameter experiences variations independently.
\textbf{2) From the perspective of training efficiency}, the state-of-the-art training methods, such as noise-injection training or selectively retraining certain layers, come with additional computational overhead compared to standard training approaches. Consequently, they result in significantly longer training time required to achieve model convergence, e.g. five times more as we shall show in Section \ref{sec: deployment efficiency}. \textbf{3) From the perspective of deployment complexity}, programming a single weight into a device with precision typically necessitates multiple read-write cycles (e.g. programming with write-verify~\cite{shim2020two}). This is done to ensure that the device's conductance value falls within the target range. Consequently, deployment time and energy consumption are significantly increased. These challenges are further exacerbated when dealing with large-scale models. 

In this work, we propose a novel approach called Tiny Shared Block (TSB) deployment, to simultaneously address the accuracy and efficiency issues caused by device variations in NVCIM DNN accelerators. TSB simply integrates a small shared $1\times1$ convolution block into the DNN architecture. Through a quick fine-tuning of TSB attached to the initially trained backbone model (e.g. early stopping), feature errors that stem from device variation-induced weight imperfection, can be wisely suppressed based on each convolution channel's importance, making the model converge to optimal solutions quickly. Our design is grounded on the following key observations: \textbf{First}, noise-injection training with an early-stop strategy can significantly save training time while providing a desired level of accuracy for further improvement. 
Since we observe that training considering the randomness of variation noises would cost a significant amount of time to progress from the initial convergence stage to full convergence, despite it can only statistically improve performance with no guarantee of optimal results for each accelerator. 
\textbf{Second}, when deploying an initially trained model (e.g. via early stopping) into an accelerator, statistically stable device variations can be captured and learned by incorporating an additional, small convolution layer (e.g. $1\times1$ convolution). The new block can be fine-tuned for suppressing spatially correlated feature errors both within and across channels based on the importance of channel features, thereby improving the accuracy, training and deployment efficiency simultaneously.




Fig. 1 illustrates our TSB method. Initially, we roughly train the backbone model to the initial convergence stage, then directly deploy it onto the crossbars. Information about weight noise on the crossbar is obtained through a read pulse following each write operation. Subsequently, a 1x1 convolution block is integrated into the backbone structure. This block is trained to learn global information, selectively enhancing features and ultimately suppressing device variation impact. Remarkably, this block is not only shared between layers but also across channels. In other words, a single tiny convolution block is shared across the entire network to surpass device variation. 
After a short training period, this block can be deployed to the same accelerator with write-verify operations. This tiny shared block can be inserted at any depth of a backbone but we recommend using this block to cover all features influenced by noise, and then the benefit of the fix function can be accumulated through the network.

Since only a small fraction of the weights require modification and write-verification on the crossbar, this diminishes both the retraining efforts and programming time. Furthermore, the relatively fixed block simplifies operational complexity. Additionally, the block utilizes the same convolution operation as the backbone, ensuring compatibility with the existing accelerator structure without the need for new functional circuits. To validate our TSB design, we have conducted extensive experiments across various backbones and datasets accompanied by hardware simulation. Simulation results confirmed the advantages of our solutions across inference accuracy, training and deployment efficiency over existing solutions. Our main contributions can be summarized as follows:
\begin{itemize}
\item We introduce a novel Tiny Shared Block method, TSB, to efficiently deploy DNN models and enhance the robustness of NVCIM DNN accelerators against device variations with small overhead.
\item The TSB method significantly improves inference accuracy on NVCIM DNN accelerators, achieving more than $20\times$ improvement over state-of-the-art (SOTA) baselines, considering the gap from the accuracy of a noise-free model.
\item The TSB method speeds up the training process by more than $5\times$, reduces weight-to-device conductance mapping time and lowers operation complexity while requiring less than 0.4\% of the original weights to be write-verified.
\item The TSB method is fully compatible with existing accelerator architectures and does not require additional functional circuits.
\end{itemize}

\vspace{-10pt}
\section{Background and Related Work}
\textbf{NVM device variation}. Non-volatile memory (NVM) devices exhibit inherent variations due to stochastic fluctuations, which include both spatial and temporal differences. This variability becomes evident when identical programming pulses are repeatedly applied to the same NVM device starting from an identical initial conductance state, often resulting in different conductance outcomes. These variations, distinct from those caused by fabrication defects, are generally independent of the physical characteristics of the device yet may be influenced by the programming targets~\cite{feinberg2018making}. Our proposed "Tiny Shared Block" (TSB) method is inherently flexible and can be adapted to account for various distributions of device variations, ensuring its applicability and effectiveness in diverse deployment scenarios of NVCIM accelerators.

\textbf{Channel attention}. The correlation between channels is often leveraged as a novel combination of features, where channel attention mechanisms highlight the varying importance of different channel features. Traditionally, several distinct layers are utilized to implement these channel attention mechanisms. Much of the existing work in this area has focused on reducing model complexity, operating under the assumption that channel relationships can be formulated by instance-agnostic functions with localized receptive fields~\cite{hu2018squeeze}. In this study, we propose a new perspective by identifying that channel correlations exhibit common characteristics when subjected to weight noise. This recognition allows us to use the common structure of channel correlations to enhance model robustness against weight noise.

\textbf{Prior Work}. In response to performance degradation in Deep Neural Networks (DNNs) due to variations in NVM devices, research has branched into two main strategies: general-level and instance-specific approaches. The general-level approach focuses on universal solutions, such as improving device reliability through advancements in materials and fabrication processes~\cite{degraeve2015causes}, and implementing write-verify operations during device programming~\cite{shim2020two}. A recent study suggests that selectively write-verifying only critical devices can significantly enhance outcomes~\cite{yan2022swim}. Additionally, strategies such as CorrectNet~\cite{eldebiky2023correctnet}, which employs modified Lipschitz constant regularization during DNN training to enhance robustness, as well as other methods involving noisy training, architectural adjustments, and pruning techniques, are also explored~\cite{yan2021uncertainty, gao2021bayesian, jin2020improving}.

Conversely, the instance-specific approach concentrates on the uniqueness of individual accelerators, often applying methods to accelerators that already have most weights programmed, with some strategies even incorporating on-chip training. Techniques such as retraining the last layer/layers~\cite{yao2020fully} use noise data from earlier layers to enhance the accuracy of final classifiers or convolution layers. 
Another method involves compensation factors~\cite{jain2019cxdnn} at the end of each crossbar column to enhance accuracy, though this requires additional functional circuits to implement.
\vspace{-10pt}
\section{TSB deployment method}\label{sec: TSB method}

Previous works on general approaches attempt to formulate the device variation, which follows a certain statistical distribution, into a mathematical model. These approaches train the network and help it adapt to the noise pattern, aiming to enhance the model's robustness. 
However, as mentioned in Section \ref{sec: intro}, such methods do not guarantee that the final convergence process—which occupies the majority of training time—will effectively improve inference performance on NVCIM accelerator instances due to the inherent randomness of weight mapping. In fact, once the weights are programmed onto crossbars, the accuracy optimization problem transitions from a non-deterministic to a deterministic domain, imposing tighter constraints on the optimization.

Instance-specific approaches are post-programming strategies that
address deterministic weight perturbations on each accelerator but are inefficient due to the large training time and resources they require, and these methods fail to leverage common features in device variation to reduce operation complexity. To overcome these limitations, we propose the TSB deployment method. This approach combines the strengths of general level methods—utilizing commonalities in device variation—and the detailed focus of instance-specific methods. The TSB method aims to reduce deployment time and operation costs while significantly improving inference accuracy.

\begin{figure}[t]
  \centering
  \includegraphics[scale=0.4]{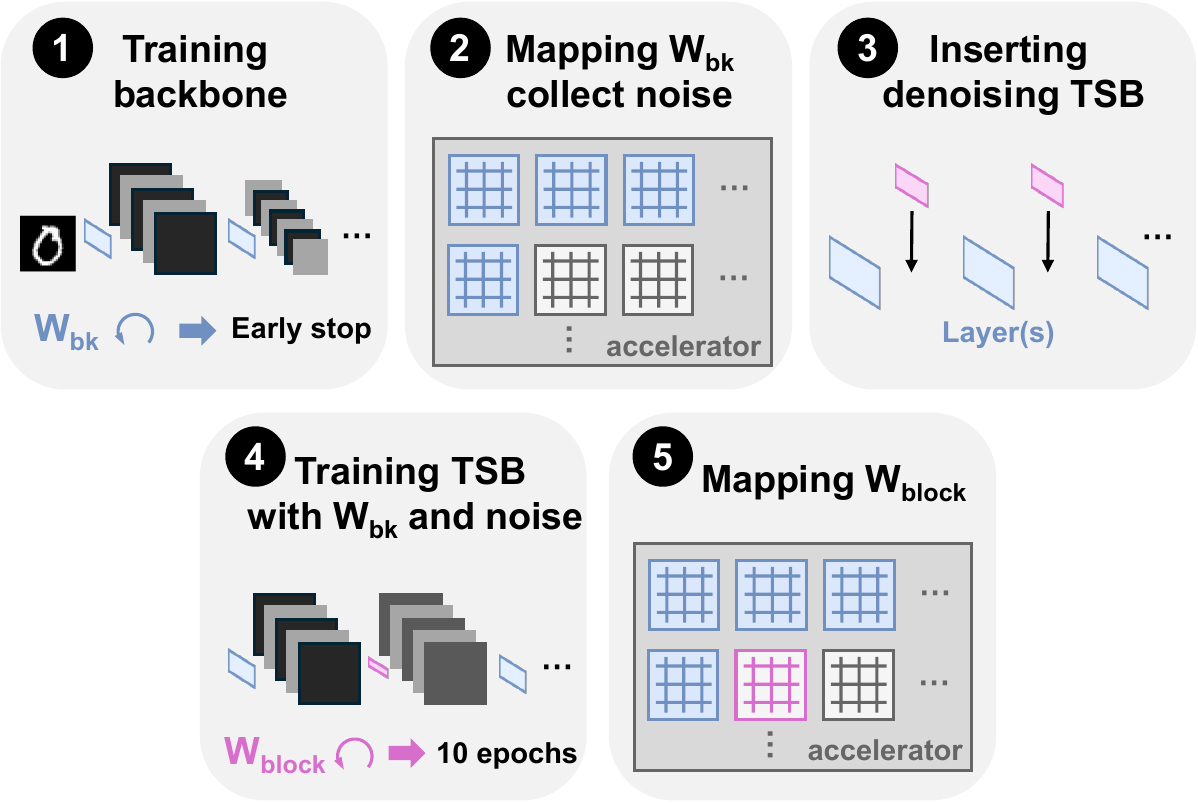}
  \caption{Workflow of TSB deployment method.}
  \Description{workflow}
  \label{fig: workflow}
  \vspace{-15pt}
\end{figure}

Fig. \ref{fig: workflow} illustrates the workflow of the TSB deployment. Initially, we train a backbone model to an early convergence stage, striking a balance between final accuracy and total training time, as this stage critically influences the final inference accuracy on the NVCIM accelerator. After deploying the backbone weights onto the crossbars, we collect noise data from these crossbars. Utilizing the trained backbone weights along with deterministic noise, we proceed to train the TSB, which is integrated into the backbone model. Following a rapid convergence training process, the TSB weights are deployed onto the crossbars using a write-verify operation. This results in a high-inference accuracy NVCIM DNN accelerator that requires small training and deployment time, as well as fewer operations only with tiny overhead. In this Section, we detail the deployment method of the TSB, outlining its structure, functionality, and the detailed process of its implementation.

\subsection{Noise Reduction Modules}\label{sec: block details}

The TSB is integrated into the DNN backbone after initial training, ensuring it does not alter the shape or form of the feature maps, thus maintaining compatibility with the original software and hardware designs. This block serves as a functional unit specifically designed to process features affected by the noise in weights. It is applicable to noisy transformations within the backbone, denoted as $\mathbf{F}_{tr}:\mathbf{X}\rightarrow\mathbf{V}, \mathbf{X}\in\mathbb{R}^{H' \times W' \times C'}, \mathbf{V}\in\mathbb{R}^{H \times W \times C}$ where $\mathbf{X}$ and $\mathbf{V}$ are feature maps. For simplicity, in the notation and experiments that follows we consider $\mathbf{F}_{tr}$ to represent convolution operations, which can be executed by either a single convolution layer or a series of layers, such as the basic or bottleneck blocks found in ResNet models. We employ the TSB as a single-layer 1x1 convolution block with same input/output channel. Based on our experimental results, optimal performance is achieved when this block is directly appended after the convolution transformation $\mathbf{F}_{tr}$ and before any non-convolution operations, such as batch normalization and nonlinear functions, for example, from [Conv layer/Conv block-BN-ReLu] to [Conv layer/Conv block-TSB-BN-ReLu].

Let $\mathbf{K} = [\mathbf{k}_{1}, \mathbf{k}_{2}, \dots, \mathbf{k}_{C}]$ denote the trained convolution kernels, where $\mathbf{k}_{i}$ represents the $i_{\text{th}}$ kernel. The transformation $\mathbf{F}_{tr}$ can be expressed as:
\vspace{-5pt}
\begin{equation}\label{eq:conv op}    \mathbf{v}_{i}=\mathbf{k}_{i}\ast\mathbf{X}=\sum_{j=1}^{C'} \mathbf{k}_{i}^{j}\ast\mathbf{x}^{j} 
\end{equation}
where $\ast$ denotes the convolution operation, $\mathbf{k}_{i}^{j}$ is a 2D spatial kernel, and $\mathbf{v}_{i}$ represents the $i_{\text{th}}$ feature map of the output features $\mathbf{V} = [\mathbf{v}_{1}, \mathbf{v}_{2}, \dots, \mathbf{v}_{C}]$. 

As the weights of kernels used for extracting spatial information are mainly concentrated around zero, the amplitude differences in noises with the same distribution tend to be relatively small. This uniform noise variably affects kernels of differing importance, leading to different impacts on feature maps according to their significance. Additionally, the noise influences the interdependency among feature channels. Consequently, the effects of noise on spatial and channel relationships are coupled, decreasing the effectiveness of simple denoising methods like scaling factors. To address this, we employ a 1x1 convolution TSB layer to decouple these relationships and make TSB effectively suppress noise across both spatial and channel dimensions simultaneously. The transformation can be expressed as:

\begin{equation}\label{eq:conv op}
\mathbf{U}=\mathbf{F}_{TSB}(\mathbf{V})
\end{equation}
where $\mathbf{U}=[\mathbf{u}_{1}, \mathbf{u}_{2}, \dots, \mathbf{u}_{C}]$ denotes the output of the block transformation $\mathbf{F}_{TSB}$. Furthermore, to ensure compatibility with original subsequent transformations, our layer maintains the input and output channel dimensions $C$ consistent. This design choice preserves the original shape and form of feature maps, ensuring that the feature map can seamlessly integrate into the network's dataflow without breaking down existing structures.

\textbf{Initialization:} The primary objective of the TSB is to improve accuracy on accelerators by leveraging the trained backbone weights and the noise data collected from the crossbars after weight mapping. To ensure that the integration of the TSB does not compromise the performance of the original backbone model, employing an appropriate initialization strategy is crucial. We utilize an \textbf{identical mapping approach} to initialize the block weights, guaranteeing that the TSB does not disrupt the pre-learned features at the onset of block training. This method is vital for preserving the completeness of the feature representations, while still permitting subsequent adjustments for de-noising. Notably, our experiments demonstrate that traditional fine-tuning weight initialization methods lead to convergence issues in this block, as they often fail to preserve the learned features from the noisy information. This underscores the critical need for our specialized initialization approach.

\subsection{Weight Sharing and Speedup}
To enhance the efficiency of the TSB by reducing both the number of parameters and the training time, we propose a novel strategy of weight sharing across layers and channels. This approach not only minimizes the parameter of TSB but also enhances the capability of the TSB to extract and process comprehensive noise information throughout the whole network. In the TSB configuration, the weights of the blocks are shared among layers. This means that any adjustments made to the block weights in one layer are automatically propagated throughout the entire network. Furthermore, within a single layer, the block weights that process different channel features are also shared. Actually, only one block is reused across the transformations in network.

Formally, for an feature set $\mathbf{V} = [\mathbf{v}_{1}, \mathbf{v}_{2}, \dots, \mathbf{v}_{C}]$, we first segment it into smaller groups $\mathbf{V} = [\mathbf{g}_{1}, \mathbf{g}_{2}, \dots, \mathbf{g}_{N}]$, where each group $\mathbf{g}_{n}$ contains $\mathbf{C}_{TSB}$ features $\mathbf{v}_{i}$, and $\mathbf{C}_{TSB}$ denoting the input and output channel capacity of the block. The number of groups, $\mathbf{N}$, is determined by:
\begin{equation}
\mathbf{N} = \left\lceil \frac{C}{C_{TSB}} \right\rceil
\end{equation}
zero padding may be required in $\mathbf{g}_{N}$ along the channel dimension to maintain consistency. The transformation applied by the TSB, $\mathbf{F}_{TSB}$, is formulated as:
\vspace{-5pt}
\begin{equation}
    \mathbf{g'}_{n}=\mathbf{g}_{n} \ast \mathbf{W}_{block}
\end{equation}
where $\mathbf{W}_{block}$ represents the shared weight of the block, and $\mathbf{g'}_{n}$ are concatenated to form the transformed feature set $\mathbf{U} = [\mathbf{u}_{1}, \mathbf{u}_{2}, \dots, \mathbf{u}_{C}]$, with padding removed after processing. This sharing ensures efficient and uniform feature processing across the network, leveraging shared resources to increase noise robustness.
This unified weight sharing, utilizing a single TSB across the entire network, dramatically reduces the number of parameters.

Due to the weight sharing between layers and channels, the weight parameters of the TSB account for only a small proportion of the total backbone weight—specifically, 0.47\%, 0.13\%, and 0.15\% for LeNet-3, VGG-8, and ResNet-18, respectively, in our experiments. 
Consequently, during the block training process, because only a very small number of weights' gradients need to be calculated, the training time for each iteration is significantly reduced.
Moreover, also because of the small size of the trained weights, the model converges much faster than conventional training methods (usually in no more than 10 epochs).
Considering the early stop of the backbone training (see Section \ref{sec: deployment process}), the overall training process is significantly accelerated. 

Additionally, as our method requires only write-verify operations on TSB weights, the time needed for these operations is greatly reduced compared with the methods in which entire backbone weights need write-verify. Overall, with the reduced computational load, \ydel{and} rapid convergence in training, and few write-verify operations, the TSB method achieves efficient deployment.

\vspace{-10pt}
\subsection{Deployment Process}
\label{sec: deployment process}

The deployment of our methodology within a NVCIM accelerator is primarily aimed at enhancing noise robustness and inference accuracy. This Section provides a detailed description of the step-by-step deployment process as outlined in Algorithm \ref{deployment_algo}. In all, this method includes two training stages.

The backbone model is initially trained with a general-level NVCIM training method to develop a noise-tolerant model. Since our method is intended to be a universal solution applicable across various networks and devices. No specific CIM training method is preferred in the algorithm. However, this universality allows for the adaptation of various established strategies from previous research to optimize model training for CIM applications.

Training stops when the model reaches a target convergence stage, which demonstrates sufficient noise tolerance for the model. The metric for stopping the training is suggested as simulated inference with noise accuracy, reflecting how the backbone will perform when deployed on accelerators. The chosen early stop point for the training process is crucial as it directly impacts the final inference accuracy and the total training time after the integration of TSB. This early stop offers flexibility, allowing users to adjust based on their specific needs. However, it is important to note that this flexibility involves a trade-off between time, accuracy, and energy consumption.

When backbone weights are loaded onto an accelerator, the deterministic noise data for each weight can be acquired at little additional cost by following a read signal immediately after each write signal. Subsequently, the modified model, which is inserted by TSB, loads the backbone weights along with their noise data and focuses on training and updating the block's weights. This method enables the block to adaptively learn the accelerator's characteristics based on instance-specific noise, thereby enhancing accelerator performance. Since this training process acts like fine-tuning, it requires a smaller learning rate than previously used, yet it reaches the same inference accuracy with a substantially smaller training time compared to traditional methods. Additionally, TSB training can also utilize the general level method, allowing TSB weights to adapt to device variations. This makes it possible for the deployment of TSB weights onto the crossbars that are homogeneous with the backbone weights, thus simplifying the accelerator's structure.

After training, the weights of TSB are deployed onto the crossbars with the same devices as the backbone weights based. However, since these weights act as corrections to the features and TSB is reused throughout the network, the weights are sensitive to device variation. Therefore, the write-verify operation is implemented to ensure that the variation of devices, on which TSB weights are deployed, falls within the target range. Fortunately, these weights take only a small portion of the total backbone weights, making our approach more time-efficient compared to methods that require write-verify operations for all backbone weights. Moreover, the complexity of the circuit required by the write-verify operation is significantly reduced due to the relatively fixed location of the block weights. Additionally, the independence of the block weights ensures that the write-verify operation is physically isolated from the backbone, thus avoiding the impact of additional operations on the weight data.

\vspace{-5pt}
\begin{algorithm}
    \renewcommand{\algorithmicrequire}{\textbf{Input:}}
	\renewcommand{\algorithmicensure}{\textbf{Output:}}
    \caption{Deployment details}
    \label{deployment_algo}
    \begin{algorithmic}[1]
    \REQUIRE DNN backbone topology $\mathcal{M}$, backbone weights $\mathbf{W}_{bk}$, block weights $\mathbf{W}_{block}$, number of first and second training epochs $ep_{1}, ep_{2}$, dataset $\mathcal{D}$;
    \FOR{($i=0$; $i < ep_{1}$; $i++$)}
        \STATE Train backbone model $\mathcal{M}(\mathcal{D}, \mathbf{W}_{bk})$ using general level CIM training methods;
        \IF{required convergence stage is reached}
            \STATE Break;
        \ENDIF
    \ENDFOR
    \STATE Deploy $\mathbf{W}_{bk}$ to the NVCIM accelerator and collect noise data $\mathbf{noise}$;
    \STATE Reconstruct model with TSB using identical mapping initialization;
    \FOR{($i=0$; $i < ep_{2}$; $i++$)}
        \STATE Train TSB with $\mathcal{M}(\mathcal{D}, \mathbf{W}_{bk}, \mathbf{W}_{block}, \mathbf{noise})$ using general level CIM training methods;
    \ENDFOR
    \STATE Deploy $\mathbf{W}_{block}$ with write-verify operation;
    \ENSURE High-performance DNN NVCIM accelerator, characterized by enhanced noise robustness and improved inference accuracy.
    \end{algorithmic}
\end{algorithm}

\vspace{-15pt}
\section{Experiments}
\label{sec: Experiments}

In this Section, we provide experimental evidence to demonstrate the efficiency and effectiveness of the TSB deployment method in reducing training and deployment time, as well as in enhancing inference accuracy. It should be noted that in this study, the TSB was inserted only after the backbone convolution layers. This decision was based on the following considerations and observations: Compared to convolution operations, the operational complexity of fully connected layers is relatively low. Moreover, the feature relationships within fully connected layers are simpler. Consequently, training methods in previous works, such as the noise-injection method, are sufficient to address the effects of noise on these layers.

We begin by discussing our noise model, which explains the relationship between device variation and the noise introduced to the weights during training. We then detail the TSB implementation in our experiments and compare our method with state-of-the-art deployment baselines across various models and datasets, focusing on total training time, number of weight modifications, deployment time, and inference accuracy improvement. Our results show significant enhancements in total training time, reduction in deployment operation time and complexity, and improvement in NVCIM performance. At the same time, we present two alternative design methods for the related accelerator design and hardware simulations, which provide practical proof of performance advantages for our method.

\vspace{-10pt}
\subsection{Deice Variation Model}
\label{sec: variation model section}

Without loss of generality, we primarily focus on device variations originating from the programming process of NVM devices, where the actual programmed conductance value deviates from the targeted value. In the following discussion, we will demonstrate how to model these device variations and measure their influence on network weights and accelerator simulations.

Set the desired weight precision of a DNN model in an NVCIM accelerator to be $M$ bits. The quantized weight $\mathbf{W}_{quant}$ is then given by the following equation:
\vspace{-5pt}
\begin{equation}
    \mathbf{W}_{quant} = \frac{\max{\left |\mathbf{W}\right |}}{2^M - 1}\sum_{i=0}^{M-1}{m_i \times 2^i}
\end{equation}
where $\mathbf{W}$ represents the original weights of the DNN, 
$\max\left |\mathbf{W}\right |$ is the maximum absolute value of the weights, ensuring the scaling factor normalizes the weights into the dynamic range determined by $M$. Each $m_{i}\in\{0,1\}$ represents the binary coefficients derived from the quantization process, mapping the floating-point weights into a fixed-point format that fits within the specified bit precision.
For an NVM device capable of representing $K$ bits of data, each weight can be stored across $M/K$ devices\footnote{assuming $M$ is a multiple of $K$ for simplicity.}, The mapping process for converting a quantized weight to its corresponding device conductance is given by the following equation:
\vspace{-5pt}
\begin{equation}
    g'_{j} = \sum_{i=0}^{K-1} m_{j \times K + i} \times 2^i
\end{equation}
where $g'_{j}$ is the conductance value stored in the $j_\text{th}$ NVM device, and $m_{j \times K + i}$ represents the 
$i_\text{th}$ bit of the $j_\text{th}$ segment of the weight. Note that negative weights can be mapped in the same manner to a separate crossbar array, allowing for a symmetrical representation of both positive and negative values. Considering device variation, the actual device conductance after programming is denoted by $g_{j} = g'_{j} + \Delta{g}$ and $\Delta g$ represents the deviation from the desired conductance value and is assumed to follow a Gaussian distribution. Thus, the actual weight $\mathbf{W}_{real}$ represented by programmed NVM devices is:
\vspace{-5pt}
\begin{equation}
    \mathbf{W}_{real} = \mathbf{W}_{quant} + \frac{\max |\mathbf{W}|}{2^M - 1}\sum_{j=0}^{M/K-1} \Delta g\times 2^{j\times K}
\end{equation}

We adopted parameter settings consistent with previous studies~\cite{jiang2020device, yan2022swim}. Specifically, we set $K$ to 2, indicating 2-bit precision for a single device conductance. For the precision of a single DNN weight, we selected $M$ to correspond to 8-bit precision. Regarding device variation, we modeled it using a Gaussian distribution with $\Delta g \sim \mathcal{N}(0, \sigma_d^2)$~\cite{qin2023negative}, where $\sigma_d$ denotes the relative standard deviation of the conductance relative to the maximum conductance achievable by a single device. We applied different constraints on $\sigma_d$: $\sigma_d \leq 0.2$ for devices with backbone weights and a more tight constraint $\sigma_d \leq 0.004$ for write-verified devices associated with block weights~\cite{shim2020two}.

Due to the random nature of sampling noise instances, a Monte Carlo simulation with noise-injection inference was conducted to accurately evaluate the actual performance of the method on the NVCIM accelerators. In each simulation run, noise was independently sampled from the target Gaussian distribution, and the average accuracy was calculated across these runs. It is important to note that when evaluating TSB with noise injection inference, the noise instances for the backbone weights are kept deterministic. This setup simulates the real-world scenario where the weights are programmed onto the accelerator.

\vspace{-10pt}
\subsection{Experimental Setup}
\label{sec: setup}

We conducted network training and simulation experiments using three different network architectures across datasets: LeNet-3~\cite{lecun1998gradient} on MNIST, VGG-8~\cite{simonyan2014very} on CIFAR-10, and ResNet-18~\cite{he2016deep} on Tiny ImageNet. The LeNet-3 network consists of two convolutional layers with output channels of 5 and 10, followed by one fully-connected layer. The VGG-8 network is composed of six convolutional layers, with successive pairs of layers having output channels of 128, 256, and 512, and includes two fully-connected layers. ResNet-18 is configured according to its standard specifications. To balance training speed and accuracy, we configured the input/output channels of the TSB at 5, 128, and 128 for these three models, respectively. The block is inserted after each convolutional layer in LeNet-3 and VGG-8, and after each basic block in ResNet-18. Initial training learning rates were set at 1e-3 for LeNet-3 and VGG-8, and 1e-2 for ResNet-18, while block training learning rates were set at 1e-3 for LeNet-3, 1e-4 for VGG-8, and 1e-3 for ResNet-18. TSB weights are supposed to be deployed on devices with $\sigma_{d}=0.004$ after write-verify operations.
Other training hyperparameters, such as batch size and learning rate schedulers, follow best practices for training noise-free models.

Experiments were conducted with PyTorch package and NVIDIA TITAN XP GPUs. To mitigate the effects of randomness and ensure reproducibility, all experimental results represent the average of at least five independent runs, unless otherwise specified. The evaluation of average inference accuracy was based on Monte Carlo simulation with noise injection inferences for 200 independent runs each. Furthermore, all statistical results are presented with a 95\% confidence interval, featuring a perturbation range of ±0.01, in line with the central limit theorem.

\vspace{-10pt}
\subsection{Implementation Details}

\begin{figure*}[ht]
  \centering
  \includegraphics[scale=0.58]{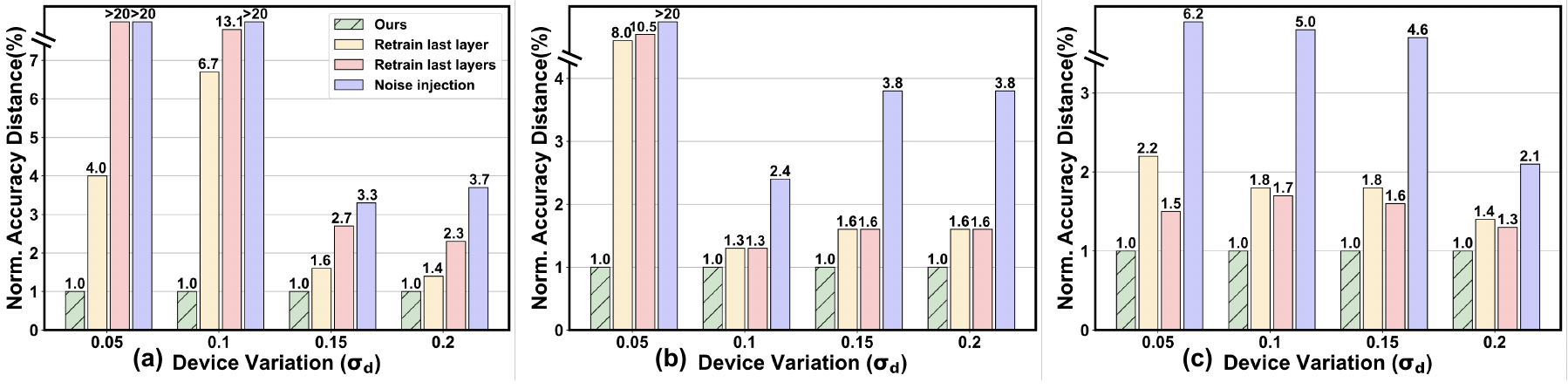}
  \vspace{-10pt}
  \caption{Normalized accuracy distances (the lower the batter) between inference accuracy and the ideal accuracy across different backbone device variation $\sigma_{d}$: (a) LeNet-3 on MNIST, (b) VGG-8 on CIFAR-10, and (c) ResNet-18 on Tiny ImageNet.}
  \Description{accuracy distance}
  \label{fig: accuracy distance}
  \vspace{-10pt}
\end{figure*}

\begin{algorithm}
    \renewcommand{\algorithmicrequire}{\textbf{Input:}}
	\renewcommand{\algorithmicensure}{\textbf{Output:}}
    \caption{Implementation example}
    \label{implementation example}
    \begin{algorithmic}[1]
    \REQUIRE backbone topology $\mathcal{M}$, weights $\mathbf{W}_{bk}$, TSB weights $\mathbf{W}_{block}$, training epochs $ep_{1}, ep_{2}$, learning rates $\eta_{1}, \eta_{2}$, dataset $\mathcal{D}$, noise distribution $\mathbf{Dist}_{1},\mathbf{Dist}_{2}$;
    \FOR{($i=0$; $i < ep_{1}$; $i++$)}
        \STATE Sample $\Delta\mathbf{W}_i$ from $\mathbf{Dist}_{1}$;
        \FOR{$x$, $\hat{y}$ in $\mathcal{D}$}
            \STATE $Loss$ = criterion($x, \hat{y}, \mathbf{W}_{bk} + \Delta\mathbf{W}_i$);
            \STATE $\mathbf{W}_{bk} := \mathbf{W}_{bk} - \eta_{1} \cdot \frac{\partial Loss}{\partial (\mathbf{W}_{bk} + \Delta\mathbf{W}_i)}$;
            \IF{required convergence is reached}
            \STATE Break;
            \ENDIF
        \ENDFOR
    \ENDFOR
    \STATE Deploy $\mathbf{W}_{bk}$ on crossbars of an NVCIM accelerator;
    \STATE Collect crossbar noise data $\mathbf{noise}$;
    \STATE Modify $\mathcal{M}$ with TSB, $\mathcal{M'}$;
    \FOR{($i=0$; $i < ep_{2}$; $i++$)}
        \STATE Sample $\Delta\mathbf{W'}_i$ from $\mathbf{Dist}_{2}$;
        \FOR{$x$, $\hat{y}$ in $\mathcal{D}$}
            \STATE $Loss$ = criterion($x, \hat{y}, \mathbf{W}_{bk} + \mathbf{noise}, \mathbf{W}_{block} + \Delta\mathbf{W'}_i$);
            \STATE $\mathbf{W}_{block} := \mathbf{W}_{block} - \eta_{2} \cdot \frac{\partial Loss}{\partial (\mathbf{W}_{block} + \Delta\mathbf{W'}_i)}$;
        \ENDFOR
    \ENDFOR
    \STATE Deploy $\mathbf{W}_{block}$ on the accelerator with write-verify operation;
    \ENSURE High inference accuracy NVCIM accelerator.
    \end{algorithmic}
\end{algorithm}

Here we utilize the widely adapted CIM general method noise injection algorithm to detail the TSB deployment method in our experiments, as shown in Algorithm \ref{implementation example}.

During both the initial training stage and the block training stage, we employ the noise-injection method to enhance weight robustness against device variation. In the initial backbone training, we sample a variation instance from a distribution $\mathbf{Dist}_{1}$, which corresponds to the target accelerator's device variations. This sampled variation $\Delta \mathbf{W}_{i}$ is added to the weights $\mathbf{W}_{bk}$ during the feed-forward process. The $\mathbf{W}_{bk}$ and $\Delta \mathbf{W}_{i}$ are then used in back-propagation to compute the gradient, after which the variation-free weights $\mathbf{W}_{bk}$ are updated using the gradient descent method. This initial training phase ends once a specific convergence metric is met.

After deploying the backbone weights $\mathbf{W}_{bk}$ on the NVCIM accelerator and collecting noise data, the TSB is inserted following each convolution transformation and initialized by an identical mapping method. The new noise instance $\Delta \mathbf{W'}_{i}$ is sampled from another distribution $\mathbf{Dist}_{2}$—specific to the write-verified device variation—is used to only update the block weights $\mathbf{W}_{block}$ by noise-injection. The learning rate for this stage may be lower than that of the initial training process (e.g., $\eta_{2}=0.1 \times \eta_{1}$).

TSB weights are then deployed on the accelerator with write-verify operations. For the deployment of TSB weights, various designs can be adopted: one could use a common crossbar to implement the TSB for improved area and energy efficiency, or duplicate the weights and avoid reuse scheduling to enhance latency.

\vspace{-5pt}
\subsection{Accuracy Improvement}
\label{sec: accuracy}

Utilizing the noise on backbone weights that have been programmed into the crossbars can effectively improve the final inference accuracy. We compare the performance of the TSB method against three baseline approaches:
\begin{enumerate}
    \item \textbf{Gaussian Noise-Injection Training~\cite{yan2022swim}:} The backbone network is initially trained via noise-injection and then directly deployed on the crossbars.
    
    \item \textbf{Retraining the Last Layer~\cite{yao2020fully}:} After deployment of the fully-converged backbone network on the crossbars, the last classifier layer is retrained using either noise-injection training or on-chip training.
    
    \item \textbf{Retraining Last Layers~\cite{yao2020fully}:}
    Similar to (2), but for networks:
    \begin{itemize}
        \item \text{LeNet-3:} The last fully connected layer and the last convolution layer are retrained.
        \item \text{VGG-8:} The last two fully connected layers are retrained.
        \item \text{ResNet-18:} The last fully connected layer and the last basic block layers are retrained.
    \end{itemize}
\end{enumerate}

In our evaluation, certain orthogonal methods, such as selective write-verify, negative feedback training, or compensation factors, were not included in the comparison since the TSB method can be employed in conjunction with these techniques.

Fig.~\ref{fig: accuracy distance} illustrates the normalized accuracy distances (the lower the better) from various methods with different backbone device variations (TSB device variation set as $\sigma_{d}=0.004$, as suggested in Section \ref{sec: setup}). Initially, we train the backbone model and deploy it to crossbars. To ensure a fair comparison, we train each backbone to its optimal convergence stage without early stop. Subsequently, we enhance the inference accuracy using different techniques and calculate the distance between the improved accuracy and the ideal accuracy. The ideal accuracy, generated by the noise-free model, is 98.9\% for LeNet, 92.97\% for VGG-8, and 61.59\% for ResNet-18.
Our TSB method significantly outperforms all baselines in different device variations. The TSB method shortens the accuracy distance more than $20 \times$ for the LeNet-3 and VGG-8 models, and $6.2 \times$ for the ResNet-18 model. Even when compared to the best-performing baseline, the improvement rates are substantial, achieving $6.7 \times$, $8.0 \times$, and $1.7 \times$ for LeNet-3, VGG-8, and ResNet-18, respectively.

The results demonstrate that our TSB method significantly enhances the inference accuracy of NVCIM accelerators facing device variation. This improvement confirms the vital role of the block structure in actively decoupling and suppressing noise within feature maps.


\vspace{-10pt}
\subsection{Deployment Efficiency}
\label{sec: deployment efficiency}

\begin{figure}[b]
\vspace{-15pt}
  \centering
  \includegraphics[scale=0.4]{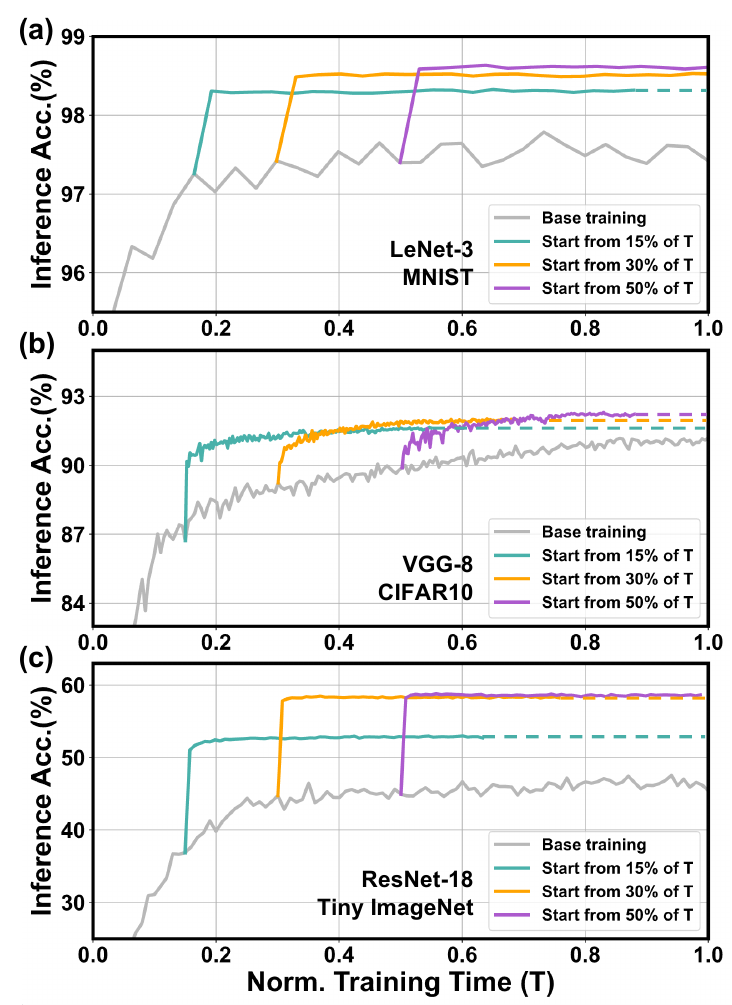}
  \caption{Convergence process of the VGG-8 model with and without the TSB method applied.}
  \Description{mix training}
  \label{fig: mix training}
\end{figure}

In this Section, we demonstrate the efficiency of the TSB deployment method. Figure \ref{fig: mix training} illustrates the convergence process of the VGG-8 backbone model with and without the TSB method. For all training and inference, the backbone device variation was set at $\sigma_{d}=0.1$, while TSB device variation was maintained at $\sigma_{d}=0.004$.

In Fig. \ref{fig: mix training}, the basic training (gray line) outlines the convergence process of the backbone through noise-injection training. We define the complete convergence time $T$ as the normalized unit for comparative analysis. Since other baselines, described in Section \ref{sec: accuracy}, rely on fully converged backbone models, their training times are not less than $T$, not to mention the additional time required for retraining.

The colored lines indicate experiments where we implement early stops at 15\%, 30\%, and 50\% of the total backbone training time. Subsequently, we conduct TSB training using these backbones and their noise on crossbars. The results demonstrate that models employing the TSB deployment method not only achieve higher accuracy than the baseline but also require less training time. For instance, the whole process of the TSB method—when it starts from the 15\% of $T$ early stop checkpoint (green line)—requires only 19.24\%, 22.16\%, and 21.45\% of $T$ training time for LeNet-3, VGG-8, and ResNet-18, respectively, to reach a converged high-accuracy model. It is observed that extending the duration of backbone training enhances the accuracy of the final model up to a saturation point, but as we discussed, there is a trade-off between the final accuracy and the training time.

In this work, the TSB training process converges within 10 epochs and the examples presented in Fig. \ref{fig: mix training} are not representative of the optimal training setting. Consequently, we can claim that the TSB deployment method requires less than 20\% of the total training time to achieve the same accuracy performance as the noise-injection baseline, thereby accelerating the training process by more than 5 times.

In addition, we observed that the proportion of TSB weights relative to the total backbone weights is tiny, taking only 0.47\%, 0.13\%, and 0.15\% for LeNet-3, VGG-8, and ResNet-18, respectively. 

Thus, compared with baselines, this TSB deployment method not only makes the gradient calculation during training more efficient but also simplifies the write-verify process during deployment. Such efficiency underscores the potential of the TSB training for edge GPU training, where resources are notably constrained, and indicates the small overhead of the NVCIM accelerators.

\vspace{-5pt}
\subsection{Overhead Simulation}

Due to the unique design that involves TSB weights being shared or reused multiple times, we have developed two distinct designs for NVCIM accelerators to accommodate this feature, which we have simulated using the NeuroSim~\cite{peng2019dnn+} platform. Specifically, one design involves mapping TSB to a public crossbar in the accelerator, allowing this part to be reused across the entire network (\textbf{C}ommon), so energy consumption can be reduced. Alternatively, we can duplicate the weights of TSB across different crossbars to enhance parallelism and decrease latency (\textbf{S}eparate). Table \ref{tab: neurosim} shows the NeuroSim simulation of base accelerators and TSB accelerators with different designs, all simulation bases on 2-bit RRAM cell and 8-bit weight and other default settings.

Compared with the original accelerator design, our designs show $0.07\times$ (C) and $0.09\times$ (S) for LeNet-3, $0.86\times$ (C) and $0.69\times$ (S) for VGG-8, and $0.41\times$ (C) and $0.48\times$ (S) for ResNet-18 overhead on EDP (Energy-Delay Product) evaluation. Considering the improvements our TSB deployment method offers in training time, write-verify operations, and inference accuracy, we believe that the associated overhead is both reasonable and acceptable.


\begin{table}[h]
\caption{Simulation for accelerators with different designs}
\label{tab: neurosim}
\begin{tabular}{lrrr}
\toprule
Model        & Energy ($\mu J$) & Latency ($\mu s$) & EDP ($\mu J\cdot\mu s$) \\ 
\hline
\midrule
LeNet-3      &  0.52    &  34.91     &   18.10   \\
LeNet-3(C)   &  0.53    &  36.03     &   19.35   \\
LeNet-3(S)   &  0.55    &  35.96     &   19.81   \\ 
\midrule
VGG-8        & 17.58    & 128.74     & 2.26e+3   \\
VGG-8(C)     & 19.44    & 216.58     & 4.21e+3   \\
VGG-8(S)     & 21.31    & 179.07     & 3.82e+3   \\ 
\midrule
ResNet-18    & 21.44    &  67.79     & 1.45e+3   \\
ResNet-18(C) & 22.25    &  92.04     & 2.05e+3   \\
ResNet-18(S) & 23.96    &  89.65     & 2.15e+3   \\ 
\bottomrule
\end{tabular}
\end{table}

\vspace{-15pt}
\subsection{Ablation Study}
In this Section, we present ablation studies for the TSB method. The example experiments use VGG-8 on the CIFAR-10 dataset, incorporating device variation with $\sigma_d=0.1$.

\textbf{TSB channels:}
Table 2 illustrates the impact of varying TSB channel numbers on the inference accuracy and training time per epoch. Notably, due to the effects of TSB reuse and feature padding operations, increases in the number of channels do not always result in a linear mapping in inference accuracy and training time. 
Specifically, when the channel number of TSB is smaller than that of the input features, reuse of TSB will lead to an increase in processing time. Conversely, when the TSB channel count is relatively large, the information learned by TSB becomes dispersed among TSB weights, resulting in the inability of smaller channel features to fully utilize the TSB information.
There is a trade-off between accuracy and training time, and the optimal setting may appear around 128 for VGG-8. Consequently, we have determined that using 128 channels in our experiments offers the optimal balance between accuracy and time efficiency.

\begin{table}[h]
\caption{TSB channel number selection}
\label{tab: channel}
\begin{tabular}{lrrrrrr}
\toprule
Channels      & 16    & 32    & 64    & 128   & 256   & 512   \\ 
\midrule
Accuracy(\%)  & 91.96 & 92.08 & 92.04 & 92.12 & 92.01 & 91.55 \\ 
Norm. Time & 1.43  & 1.19  & 1.07  & 1.00 & 1.11  & 1.30  \\ 
\bottomrule
\end{tabular}
\end{table}
\vspace{-10pt}

\textbf{\# of TSB transformation:} This part demonstrates the effect of inserting a varying number of TSB transformation layers into the backbone on inference accuracy. We experiment with inserting 1 to 6 TSB layers into the backbone respectively, with all backbones maintain in the same fully converged state. The experiments yielded inference accuracies of 91.96\%, 92.02\%, 91.91\%, 92.05\%, 92.06\%, and 92.12\% for the 1 to 6 TSB transformation layers, respectively. The results indicate that inference accuracy on the NCVCIM accelerator improves as the number of TSB transformation increases. This enhancement is primarily attributed to the TSB's ability to adaptively learn noise patterns through weight sharing. Therefore, incorporating more TSB transformations in the backbone enhances exposure to noise, contributing to more effective learning and subsequently higher inference accuracy."

\textbf{Different layers within TSB:} 
We use different numbers of 1x1 convolution layers to construct TSB for DNN deployment. Specifically, we implement TSB with 1 to 4 convolution layers. Operating with the same fully-converged backbone model, the respective inference accuracies achieved by the TSB method are 92.12\%, 91.81\%, 91.73\%, and 91.53\%. Concurrently, the training time for each epoch increases with the number of layers. Notably, the TSB with a single layer exhibits the best performance. This results from that multi-layer TSB introduces additional correlations to features, which hinders the decoupling process of noise and features.
\section{Conclusion}
\label{sec: Conclusion}
In this work, we introduced the Tiny Shared Block (TSB) deployment method as an efficient solution for deploying DNNs on NVCIM accelerators. The application of this method across various models demonstrates significant advantages, including reduced training time, simplified deployment complexity, and enhanced inference accuracy, with small overhead. Given these benefits, the TSB deployment method is expected to offer a promising solution for addressing the challenges associated with deploying DNN models on NVCIM accelerators.

\begin{acks}
This research was supported by ACCESS (AI Chip Center for Emerging Smart Systems), sponsored by InnoHK funding, Hong Kong SAR.
\end{acks}

\bibliographystyle{ACM-Reference-Format}
\bibliography{Ref}

\end{document}